\documentclass[12pt,letterpaper]{article}
\usepackage[utf8]{inputenc}
\usepackage[english]{babel}
\usepackage{amsmath}
\usepackage{amsfonts}
\usepackage{amssymb}
\usepackage{graphicx} 
\usepackage{listings}
\usepackage{color}
\usepackage{float}
\usepackage[left=2cm,right=2cm,top=2cm,bottom=2cm]{geometry}
\usepackage[hidelinks, pdflang={en-US}, pdftitle={Paper}]{hyperref}
\usepackage{xcolor}
\usepackage[separate-uncertainty = true,multi-part-units=single]{siunitx}
\DeclareSIUnit\year{yr}

\allowdisplaybreaks

\newcommand{\authorstyle}[1]{{\large\usefont{OT1}{phv}{b}{n}\color{black}#1}} 

\newcommand{\institution}[1]{{\footnotesize\usefont{OT1}{phv}{m}{sl}\color{black}#1}} 

\usepackage{titling} 

\newcommand{\HorRule}{\color{gray}\rule{\linewidth}{1pt}} 

\pretitle{
	\vspace{-30pt} 
	\HorRule\vspace{10pt} 
	\fontsize{26}{30}\usefont{OT1}{phv}{b}{n}\selectfont 
	\color{black} 
}

\posttitle{\par\vskip 15pt} 

\preauthor{} 

\postauthor{ 
	\vspace{-10pt} 
	\par\HorRule 
	\vspace{20pt} 
}

\title{A Social Force Model of the Evacuation from a Big Box Store}

\author{
	\authorstyle{Gavin Buxton} 
	\newline\newline 
	\institution{Science Department, Robert Morris University, Moon Township, PA 15108, US.}\\ 
}

\date{}

\begin{document}

\maketitle 

\vspace{-80pt}
  
We include elliptical cross-sections to physically represent people, and irregular polygons to represent wheelchair users, in an anisotropic social force model whose velocity and angular dependence also captures the social tendency for people to avoid walking into one another. 
Physical interactions are included that depend on the area of overlap between people, or obstacles, to capture normal forces that resist compression and tangential forces that resist sliding motion.
The model is further extended to include decision making capabilities, small social groups, the spread of panic, and herding behavior.
A large box store is simulated during an evacuation where people move through the store, along the shortest path, to their desired exits.
The effects of exit choice, or the perceived availability of exits, on exit times is elucidated.
It is found that ignoring `staff only' exits, and only exiting from the main entrances, can significantly increase average egress times.

\section*{Introduction}

Understanding crowd dynamics in emergency situations can optimize evacuation strategies for both speed and safety, and potentially reduce casualties.
For example, during structural fires or terrorist attacks evacuees may be subjected to smoke-filled environments and some exits may be unavailable \cite{cao2023development, lu2023crowd}.
During active shooter situations it is important to understand how run, hide, and fight tactics (and other factors, such as assault weapons bans) may influence the severity of increasingly frequent mass shootings \cite{hayes2014agent, anklam2014mitigating, briggs2016active, glass2018use, arteaga2020building, arteaga2023effect, lavalle2023effectiveness}.
Other disasters, such as shipwrecks \cite{kang2019improved} or earthquakes may also offer unique difficulties for crowds trying to evacuate.
Even in the absence of disaster scenarios, people may be crushed or trampled as crowds stampede during large events and sometimes with no apparent cause \cite{helbing2000simulating, sticco2021social, barr2024beyond}.
Experimentally elucidating the dynamics of panicked crowds in real emergency situations is obviously unethical \cite{helbing2000simulating, cornes2021microscopic}.
However, the effects of individual characteristics and capabilities, and the consequences of these on the efficiency and safety of evacuation dynamics, can be efficiently predicted from computer simulations \cite{helbing1995social}.

The social force model of pedestrian dynamics is an example of agent based modeling (ABM) which captures the dynamics of individuals subject to both physical and social forces \cite{helbing1995social}.
The physical forces may arise from interactions between an individual and other people or objects, while the social forces capture internal motivations such as a desire to move in a given direction at a given speed, the desire for personal space or an attraction towards friends and family \cite{helbing1995social}.
The evacuation in primarily captured with an acceleration towards an assigned speed in the direction towards the persons destination \cite{helbing1995social, helbing2005self}.
People may feel uncomfortable moving in close proximity to others or near obstacles, and a repulsive interaction that captures the respect of personal space and a desire to avoid collisions is included that often takes into consideration the direction of motion and the tendency for individuals to pass on a particular side \cite{helbing1995social, helbing2005self, moussaid2009experimental}.
Physical (or granular) forces may also arise, especially at high densities, as people in contact with each other become compressed, or experience friction as they attempt to slide past one another \cite{helbing2005self, moussaid2009experimental, sticco2020effects}.
Similar granular forces are also often considered between people and walls, or other obstacles \cite{helbing2000simulating, helbing2005self}, and attractive forces may also exist between a person and their family and friends \cite{helbing1995social}.
Through these simple interactions a crowd may be predicted to form lanes consisting of pedestrians walking in opposite directions \cite{helbing1995social, moussaid2012traffic, dong2019state}, people alternating between which side passes first at bottlenecks \cite{helbing1995social, moussaid2009experimental}, or narrow exits may become clogged which can cause the `faster is slower' effect, and increased evacuation times as desired velocities increase \cite{parisi2005microscopic, helbing2005self, ha2012agent, cornes2021microscopic}.

The interaction of pedestrians can be assumed to be anisotropic.
Other people or objects in front of a person are expected to be more impactful than people or objects behind them. The socially repulsive force between a person and other people (or obstacles) can be dependent on velocity \cite{helbing1995social}. The interaction can be elliptical \cite{helbing1995social}, and possess an anisotropic factor biased in the direction of the persons velocity \cite{helbing2005self} depending on the time over which any potential collisions may occur \cite{zanlungo2011social}.
Alternatively, rather than increasing the range of repulsive interactions to avoid collisions, the desired velocity can be reduced instead \cite{smith2009modelling}.
When a person is predicted to collide with either another person or an obstacle the desired velocity of the person can be appropriately reduced \cite{smith2009modelling, lu2020pedestrian}.
The direction of the desired velocity has also been modified to avoid collisions, while still giving preference in the direction of the desired objective \cite{moussaid2011simple}.
Anisotropic interactions have also been considered in the granular forces \cite{chen2018social}, which might be more important when modeling denser, more congested, crowds.
To more accurately capture the shape of people from an aerial perspective, elliptical interactions where the semi-major axis is in the lateral direction (aligned with the shoulders of the person) have been considered instead of circular interactions \cite{johansson2007specification, fu2024application}.
Multi-circle models have also been adopted to obtain a similar effect \cite{langston2006crowd}.
In multi-circle models a center circle represents the body of the person and is supplemented with additional overlapping circles to represent the shoulders of the person \cite{langston2006crowd, smith2009modelling, chen2018social}.
The multi-circle or elliptical representation is considered more accurate at capturing the local crowd environment, and when physical interactions dominate it may be worth the additional computational expense to include these representations \cite{chen2018social}.

While most social force models represent people without disabilities, there have been attempts to capture the effects of disabled people \cite{fu2024application}.
People of different disabilities may have lower velocities on average \cite{pan2021fundamental, geoerg2023together} and take more time to change their speed or direction \cite{boyce1999toward, hashemi2018emergency}.
In the social force model, therefore, it is common to capture people with disabilities by reducing their velocity \cite{fu2024application, liu2024agent} and using larger circles \cite{liu2024agent} or larger ellipses \cite{fu2024application} to account for the larger spatial dimensions of people with crutches or wheelchairs \cite{geoerg2023together}. 

Evacuations often occur during an, at least perceived, emergency and may involve people who are panicked to various degrees.
The degree of panic that a person experiences can be influenced by many factors, such as their proximity to danger, but may also occur seemingly without cause \cite{helbing2000simulating}.
Panic can propagate like a contagion through a crowd of people, with susceptible people becoming ``infected" via communication with others that are panicking \cite{chen2018social, cornes2019fear}.
Panic may also be influenced by external factors, such as gun shots or alarms \cite{yang2014guided}.
The level of panic may increase with wait time as people become anxious when their movement is restricted \cite{helbing2000simulating}, for example, if exits are blocked or congested and the person is unable to exit and remove themselves from danger. 
As people become more panicked this causes chaotic behavior, where people respond irrationally, and this can have a detrimental effect on decision making \cite{ha2012agent, de2023large}.
For example, alternative exits may be ignored as people become more focused on their immediate objective and less aware of their surroundings \cite{helbing2000simulating, bonabeau2002agent}.
The most expected outcome to panic is an increased speed \cite{helbing2000simulating}, although when the threat is perceived to be approaching and uncontrollable, people may also freeze \cite{low2015threat}.
The social contagion isn't just limited to a persons individual behavior, however.
Herding is where panicked individuals follow others, like a herd, towards a given exit while ignoring closer or more accessible exits \cite{helbing2000simulating, bonabeau2002agent, haghani2019simulating, vermuyten2016review}.
The level of panic is often included in the social force model through a parameter than linearly increases the desired speed of people as they become more panicked or anxious \cite{helbing2000simulating}, although a nonlinear model based on the Yerkes-Dodson law has been proposed to account for impaired performance at greater levels of anxiety \cite{liu2024modified}.
Here, we consider a degree of panic that increases because of the panic of others and a person's restricted motion, but we also include a probability of switching exits to capture both panicked people forming a herd and non-panicked people rationally choosing other available exits.

People walking in crowds may belong to smaller social associations, such as colleagues, friends, couples or family groups.
Estimates vary on the number of people in crowds that might belong to a social group \cite{xu2009simulation, moussaid2010walking}, and this will depend on the nature of the crowd (e.g., the location, event and time) \cite{xu2009simulation}.
While in less dense crowds, groups of people tend to align perpendicular to the walking direction, this can change to a `v' shaped pattern at higher densities with people on the outside of the group walking ahead and the person in the center walking slightly behind the group \cite{moussaid2010walking, zanlungo2015spatial, corbetta2023physics}.
Such group organization is thought to facilitate communication between individuals \cite{moussaid2010walking}.
However, such behavior depends largely on the nature of the group \cite{zanlungo2017intrinsic, gregorj2023social} and may be different during emergency situations.
Egress times during an evacuation may increase with social groups for two reasons. First, the response to the emergency situation may be delayed in a social group, as they decide on their collective action \cite{bode2015disentangling, haghani2020empirical} and, second, they may move slower as a group \cite{bode2015disentangling}, perhaps to accommodate the slowest member of the group \cite{wagnild2013energetic}.
However, once they reach an exit there is not believed to be a delay in making their way through the exit \cite{bode2015disentangling}.
Social groups have been implemented in the social force model usually in terms of an attractive force between individuals within the group \cite{moussaid2010walking, chen2018social}.
In particular, the attractive interactions may be similar to the exponential repulsive social force interactions \cite{helbing1995social, xu2009simulation, turgut2022modeling}, or a more Hookean force term that attracts members of the group if they stray too far from the center of mass of the group \cite{moussaid2010walking, farina2017walking, corbetta2023physics}.
While isotropic interactions were initially considered \cite{helbing1995social, xu2009simulation}, more recently the field of view of an individual (or other anisotropic terms) have been considered \cite{moussaid2010walking, farina2017walking, corbetta2023physics}.
However, such anisotropic interactions may be different during emergency evacuations, with distributions of group members elongated in the direction of egress \cite{von2017empirical, xie2020evacuation}. 
Social interactions between strangers may also play a role in evacuation dynamics \cite{drury2013psychological}.
While fallen and injured people may pose a hindrance to the egress of others \cite{helbing2000simulating}, during an emergency situation they are often helped by strangers \cite{johnson1987panic, von2016modelling, cimellaro2019integrating}.
Such mutual aid behavior appears anecdotally common \cite{turgut2022modeling}.
In addition, some individuals (whether due to a natural tendency or through training) may assume a leadership role, help others by guiding them towards the most optimal exit, and ultimately improve egress times \cite{fang2016leader, cimellaro2019integrating, arteaga2023effect}.

During the evacuation process people not only decide on their course of action, but may also revisit their decisions as the situation evolves and they acquire additional information \cite{haghani2019simulating}.
The distance the person is to nearby exits is often considered a common factor when deciding which exit to leave from \cite{shi2009developing, hashemi2018emergency}.
However, the degree of congestion and crowding at exits (or at least a persons perception of potential clogging) is a time-dependent factor that may alter a persons choice of exit \cite{haghani2019simulating}.
How readily a person is willing to change their mind when faced with potential congestion may depend on how stressed and panicked the individual is \cite{bode2013human, hashemi2018emergency}.
Exit familiarity also plays a large role in the initial choice of exit \cite{shi2009developing, hashemi2018emergency}, and during an emergency evacuation from a building where the occupants are unfamiliar with alternative exits this may be the most important factor \cite{kinateder2018exit}.
When stressed and panicked, people tend to trust the exit they are most familiar with (often their point of entrance) than risk trying alternative exits, even when the familiar exit is further away and more congested \cite{sime1983affiliative, kinateder2018exit}.
This may be alleviated by increasing the visibility and signage of alternative exits \cite{zhang2017optimal}.
People also tend to follow others \cite{hashemi2018emergency, bonabeau2002agent}, especially during an evacuation when occupants are stressed and panicked \cite{bode2013human, hashemi2018emergency}, when other individuals are identified as leaders or staff \cite{hashemi2018emergency, kinateder2018exit}, or when in unfamiliar surroundings \cite{farina2017walking}.

The social force model can capture the behaviors of individual people and, through simple interactions on an individual level, complex patterns and effects can collectively emerge.
The social force model can also include and optimize different building designs, mainly through the inclusion of walls, obstacles and exits \cite{ha2012agent, arteaga2020building}.
However, to capture the evacuation from more complicated building designs with multiple exits it is necessary to accurately predict how individuals might decide on the exit they will leave from, and how these decisions might change with time \cite{zainuddin2010modification, cimellaro2019integrating, ma2021optimization}.
Here we develop a model of how a persons choice of exit, and their ability to change their mind as exits become congested or they observe others choosing alternative exits, might influence egress times. 
The model captures the elliptical shape of people, the irregular shape of wheelchair users, and the non-circular interactions of moving people in social force models.
This model, described in the next section, is applied to the evacuation dynamics from a typical big-box store (a large, box-like physical retail structure) to elucidate the effects of exit choice and perceived availability on egress times. 

\section*{Model}

The dynamics of people in the social force model starts with the following dynamic equation for capturing the acceleration of people.
\begin{equation*}
m_i \frac{d \mathbf{v}_i}{d t} = m_i \frac{ \mathbf{\hat{e}_i} v_i^0 - \mathbf{v}_i }{t_v} + \sum_{j \neq i}\mathbf{F}_{ij} + \sum_{w}\mathbf{F}_{iw} - \mathbf{F}_{i}^{fr}
\end{equation*}
where $m_i$ is the mass and $\mathbf{v}_i$ the velocity of the i\textsuperscript{th} person. $\mathbf{\hat{e}_i}$ is a unit vector that points in the direction that will lead to the shortest path to a given exit. $v_i^0$ is the desired speed of the person, and will be modified by the level of panic experienced by the individual. 
$v_i$ is the current velocity and $t_v$ is the timescale over which a person accelerates or change direction to their desired velocity.
There are three forces in the system: an interaction between neighbors, $\mathbf{F}_{ij}$, an interaction between the person and the wall, $\mathbf{F}_{iw}$, and a frictional force for wheelchair movements perpendicular to the direction they are facing, $\mathbf{F}_{i}^{fr}$. 

The layout of a big-box store is shown in Fig. 1a.  The two main entrances are doors numbered one and two. These are at the front of the store, and most people would be expected to leave through these doors, as these are also the doors they entered. Doors three and four on the left-side of the layout might lead to outdoor areas, e.g. a garden center, and may only appear accessible in the summer. Exit number five at the back of the store is close to where typically a bathroom might be located. Door six might typically be labeled as an emergency exit or staff only exit. On the right-hand side of the layout where there might be a bakery is the seventh exit, and this might only be accessible by heading through a staff only area.  However, here we consider the consequences of shoppers ignoring such restrictions in an emergency situation and potentially exiting from all exits, as opposed to being limited to just the two main entrances and exits.
Recall $\mathbf{\hat{e}_i}$ is a unit vector that will lead a person through the store, by the shortest path, to the exit that the person is heading towards.
In order to calculate $\mathbf{\hat{e}_i}$ we must calculate collision free paths from any point in the store to any exit.
Points around the corners of obstacles (or more accurately, the width of a person from the obstacles) are connected along line-of-site paths (that do not cross with walls) and the sequence of such points that result in the shortest distance to the exit that the person is heading to is calculated \cite{lozano1979algorithm}. An example of this is plotted in Fig. 1b. which shows this vector field. Note regions inside solid walls, or close enough to a solid wall that might be expected to see the person overlap with the wall, are not expected to be occupied and so the vector $\mathbf{\hat{e}_i}$ is set to the direction that will see the person no longer overlap with the wall. A stronger person-wall interaction will be described later, for when physical interactions dominate.

Note that while the social interactions in the current model are anisotropic, and sometimes referred to as an elliptical model, the people in the current model are also physically represented by ellipses as shown in Fig. 2a. The wheelchair users are represented by irregular polygons as shown in Fig. 2b. The interaction between neighbors consists of three terms in the current model. 
\begin{equation*}
\mathbf{F}_{ij} = \mathbf{F}_{ij}^s + \mathbf{F}_{ij}^c + \mathbf{F}_{ij}^g
\end{equation*}
where the three forces are due to physical collisions, $\mathbf{F}_{ij}^c$, social forces, $\mathbf{F}_{ij}^s$, and group interactions $\mathbf{F}_{ij}^g$.
The social force model employed here, which reflects a psychological tendency of people to avoid one another, is consistent with Helbing and Johansson \cite{helbing2010pedestrian}, with the exception of a rotation to mimic people passing on one side or the other. 
\begin{equation*}
\mathbf{F}_{ij}^s = \left( \lambda + (1-\lambda) \frac{1 + \mathbf{\hat{v}}_i\cdot\mathbf{\hat{d}}_{ij}}{2}\right) A_i \text{e}^{-b_{ij}/B_i} \frac{|\mathbf{d}_{ij}| + |\mathbf{d}_{ij} - \mathbf{v}_{ij}t_s|}{4 b_{ij}} \,\mathbf{R}_{\phi}\left( \frac{\mathbf{d}_{ij}}{|\mathbf{d}_{ij}|} + \frac{\mathbf{d}_{ij} - \mathbf{v}_{ij}t_s}{|\mathbf{d}_{ij} - \mathbf{v}_{ij}t_s|}\right)
\end{equation*}
The first part of this force captures the angular dependence of this interaction. People are less likely to respond to people that are behind them. This is controlled by parameter $\lambda = 0.1$, which captures a strongly anisotropic interaction \cite{helbing2010pedestrian}. The unit vectors, $\mathbf{\hat{v}}_i$ and $\mathbf{\hat{d}}_{ij}$, point in the direction of the velocity and the direction to the neighboring person, respectively.
The second part of this term consists of the elliptical specification for the interaction between two people \cite{helbing1995social, helbing2010pedestrian}. Rather than the simpler circular interaction, the elliptical specification depends on the relative velocities of the people as well as the distance between them.
$A_i$ and $B_i$ are constants that represent interaction strength and range, respectively, $\mathbf{d}_{ij}$ is the vector distance to the neighboring person, and $\mathbf{v}_{ij}$ is the relative velocity given by $\mathbf{v}_{ij} = \mathbf{v}_{i} - \mathbf{v}_{j}$; the difference between the person's velocity and that of their neighbor. $t_s$ is the time over which one might predict the locations of oneself and neighbors, given their velocities, such that the repulsive force isn't simply based on the current locations, but potential locations extrapolated over this time. 
The distance $b_{ij}$ also takes into consideration the relative velocities in the same manner and is given by
\begin{equation*}
2 b_{ij} = \sqrt{\left( |\mathbf{d}_{ij}| + |\mathbf{d}_{ij} - \mathbf{v}_{ij}t_s|\right)^2 - \left|\mathbf{v}_{ij}t_s \right|^2}
\end{equation*}
The last part of the equation is a vector quantity that dictates the direction of this repulsive force. The rotation matrix, $\mathbf{R}_{\phi}()$, is added to cause people to pass on the right, for positive angles $\phi$.
Fig. 2c. depicts the force acting on a neighbor surrounding a moving person. 
Not only is the neighbor repelled from the location of the person, but is repelled from future locations extrapolated from the velocity over the time, $t_s$. 
Note putting $t_s = 0$ would recover the traditional circular specification. 
However, in the current implementation we include a rotation matrix which cause neighbors directly in front of the person to be pushed to the left, as they pass on their right.

The collision force is calculated from the overlap of the areas occupied by the people, which in the current model are captured by two-dimensional shapes; ellipses for people without disabilities and people with crutches, and irregular polygons for people who use a wheelchair. This is depicted in Fig. 3.
\begin{equation*}
\mathbf{F}_{ij}^c = k_n A_{ij} \mathbf{\hat{n}}_{ij} + k_t A_{ij} \left(\left(\mathbf{v}_j - \mathbf{v}_i \right) \cdot \mathbf{\hat{t}}_{ij} \right)\mathbf{\hat{t}}_{ij}
\end{equation*}
The magnitude of the interaction is proportional to the area of overlap, given by $A_{ij}$. A normal force, with a force constant $k_n$, acts perpendicular to the line passing through the intersection points of the overlapping shapes. This is the force that would cause the two bodies to be pushed away from one another.
Similarly, a tangential forces, with a force constant $k_t$, acts parallel to the line passing through the intersection points and is proportional to their relative velocity in the tangential direction. In other words, if two people overlap and they are moving with different velocities such that they might slide past one another, then a frictional force would act to reduce this sliding motion.
Note the irregular shape (two rectangles) of the wheelchairs can produce two separate interactions, and when this is the case a force is simply calculated for each rectangle.

The last force acting between neighbors is a group interaction.
\begin{equation*}
\mathbf{F}_{ij}^g = k_g H(c_{ij} - d_0) \mathbf{\hat{c_{ij}}}
\end{equation*}
where $k_g$ is the force constant between people in the same group. 
Here we assume this is a constant, but in reality this might depend on the social interactions between group members (a mother might have a stronger affinity to a child, than might exist between casual acquaintances).
$H()$ is the usual Heaviside step function (equal to the argument when the argument is positive, and zero when the argument is negative).
$c_{ij}$ is the distance between a person and the ``center of mass'' of the group, which is simply defined as the average location of all people in the group. $d_0$ is a separation distance that people in a group may be comfortable with. $\mathbf{\hat{c_{ij}}}$ is a unit vector pointing from a person to the center of the group.
Note that more complicated interactions capture the social dynamics, and patterns of people walking under normal circumstances, but here we are only concerned with a more panicked response.

The next force acting on a person is the interaction between a person and a wall. 
\begin{equation*}
\mathbf{F}_{iw} =  A_w \text{e}^{-d_{iw}/B_w} \mathbf{\hat{d_{iw}}} + k_n H(A_{iw}) \mathbf{\hat{n}}_{iw} + k_t H(A_{iw}) \left(\left( - \mathbf{v}_i \right) \cdot \mathbf{\hat{t}}_{iw} \right)\mathbf{\hat{t}}_{iw} 
\end{equation*}
The first term is a soft repulsive term, similar to the social force between people, that captures the tendency of people to feel discomfort when walking close to walls or obstacles. $A_w$ and $B_w$ are constants that represent interaction strength and range, respectively. $d_{iw}$ is the distance from the person to a wall, and $\mathbf{\hat{d_{iw}}}$ is a unit vector that points away from the wall. Here, the location of the wall is the closest point along the wall to the person. 
The second and third terms represents the physical contact forces that can occur as people are pushed into walls, as shown in Fig. 3c. 
These terms are proportional to the area of overlap with the person and the wall. 
Here the wall is represented by a 45-90-45 triangle that extends back from the person to capture the \qty{90}{\degree} corners between walls; a \qty{90}{\degree} corner between two walls would consist of two \qty{45}{\degree} vertices from the triangles of both walls.
The second term is a normal force, with force constant $k_n$, that pushes the person away from the wall. This force acts perpendicular to the line passing through the intersection points of the overlapping shapes, in the direction $\mathbf{\hat{n}}_{iw}$. 
The third term is a tangential force, with force constant $k_t$, that acts parallel to the line passing through the intersection points and is proportional (but in the opposite direction) to the person's velocity in the tangential direction, $\mathbf{\hat{t}}_{iw}$. 
This can be considered the frictional force between the person and the wall.

The final force only acts on a person using a wheelchair, and is a frictional force, $\mathbf{F}_{i}^{fr}$, which is introduced in the current model to limit the perpendicular motion of wheelchairs. When the forces perpendicular to the wheelchair are less than the limit of static friction the net perpendicular force will be zero.
\begin{equation*}
\mathbf{F}_{i}^{fr} \leq \mu_s m_i g \,\mathbf{\hat{h}}_{i\perp}
\end{equation*}
In other words, up to this limit any forces perpendicular to the wheelchair will cause no motion, but once this limit is exceeded the frictional force will be $\mu_k m_i g$ in magnitude, and oppose the sideways motion of the wheelchair.

The elliptical (or irregular polygon) representation of people in the current model, as opposed to the usual circular representation, dictates that the people have orientation. Therefore, the orientation and angular velocity of people must also be updated in the current model.
\begin{equation*}
I_i \frac{d \omega_i}{d t} =  k_{\theta} ({\theta_i^0} - {\theta}_i) +  \sum_{j \neq i}r_{ij} F_{ij}^c \sin\phi_{ij} + \sum_{w}r_{iw} F_{iw} \sin\phi_{iw} - k_{\omega} \omega_i
\end{equation*}
$I_i$ is the moment of inertia, and $\omega_i$ the angular velocity, of the person.
$k_{\theta}$ is a torsional constant. $\theta_i^0$ is the desired angle that the person should be facing (obtained from the desired direction of motion, $\mathbf{\hat{e}_i}$) and $\theta_i$ is the angle the person is facing (which is not necessarily the same as the direction of their velocity). As people collide with each other, or with the surrounding walls, they can be rotated by these interactions. The second term calculates the torque from the interactions with neighboring people. $r_{ij}$ is the distance from where the forces act on the person and their center of mass, $F_{ij}^c$ is the contact forces from before, and $\phi_{ij}$ is the angle between these forces and the direction from where the forces are applied and the center of mass (assumed to be the axis of rotation). 
Similarly, in the third term $r_{iw}$ is the distance from where the forces act on the person, if they interact with a wall, and their center of mass. $F_{iw}$ is the person-wall interactions from before, and $\phi_{iw}$ is the angle between these forces and the direction from where the forces are applied and the center of mass. 
$k_{\omega} = 2\sqrt{I_i k_{\theta}}$ is a damping coefficient that ensures critical damping in the current model.
One effect of these rotations is that people can rotate sideways to slide past one another when areas are congested, however in the current model this is due to physical interactions rather than a social tendency.

In addition to the physical and social forces described above, the model also captures the panic response of people during the evacuation.
\begin{equation*}
\frac{d p_i}{d t} = k_p\left( \sum_{j \neq i} \frac{\beta_v \left(1 + \cos(\theta_{ij} - \theta_i) \right) + 2\beta_a}{2 r_{ij}^2} H(p_j-p_i)\right) + \beta_s H(f_v v_0 - v_i) - \delta_i p_i
\end{equation*}
The degree of panic is given by $p_i$ which is confined between 0 and 1; 0 representing relative calm, and 1 representing a panicked individual. $k_p$ is a constant, $\beta_v$ represents the effect of visualizing others, and is zero when the other person is not in line of sight. $\theta_{ij}$ is the angle between the person and their $j$\textsuperscript{th} neighbor and $\theta_i$ the direction the person is facing. In other words, people in the line of sight have more influence on increasing a persons panic level. $\beta_a$ represents the effect of audio communications (e.g., yelling and screaming) on increasing panic levels. 
The panic level only increases if the neighbor is more panicked and is proportional to the difference in the degree of panic between their neighbor, $p_j$, and themselves. 
These effects are assumed to be inversely proportional to the distance between people squared.
Panic is also considered to increase when a person evacuating is unable to move at their desired speed, which in the current model is represented by the rate, $\beta_s$\cite{trivedi2018agent}. In the current model we assume this occurs when the person is moving slower than some fraction of the desired velocity, given by $f_v$, and increases linearly as they move slower. $\delta_i$ is the rate at which a persons panic levels decrease. 

Herding is included in the current model through the ability for people to switch directions and follow others.
A panicked person may switch to herd-like behavior and head towards the same exit as their neighbors. The probability of such a change in heading per second, of person $i$ towards exit $d$, is given by
\begin{equation*}
P_{i, d}^{h} = \left(1 - \text{e}^{-t_c/t_x}\right) H\left(1 - \delta_{id}\frac{\rho_{d}}{\rho_{m}}\right) \left[1 -   \exp \left( -\sum_{j \neq i}\beta_{ijd} \frac{1 + \cos(\theta_{ij} - \theta_i)}{2 r_{ij}^2} \right)\right] p_i
\end{equation*}
The first part of this equation limits subsequent changes from occurring too quickly.  How long ago it was since the person last changed direction is $t_c$, and $t_x$ is a time that characterizes the time frame over which someone might change their mind in this regard. 
Inside the Heaviside step function is a term that limits changing direction to a door that is too congested, as has been observed experimentally \cite{gaire2018exit}. $\delta_{id}$ is equal to one if the door is in the line of site of the person and equal to zero if the door is not visible. $\rho_d$ is the density of people at the exit (within \qty{2}{\m} of the door), and $\rho_m = 4\,\text{people/m}^2$ is the maximum density of people before injuries might typically occur \cite{helbing2010pedestrian, turgut2022modeling}. 
Inside the square parenthesis, the one minus exponential term and the inverse square of the distance provides a nice transition from people being readily influenced by neighbors close by, but not at all by neighbors further away. 
The range of this interaction is controlled by $\beta_{ijd}$, which is assumed to be zero if the neighbor is not in the line of site of the person.
Only neighbors in front of the person can influence their decision to change direction, or when the angle between the person and their neighbor, $\theta_{ij}$, is close to the angle the person is facing, $\theta_i$.
The last term is the current panic level of the person, as herding is associated with panicked individuals. The more panicked a person becomes the more likely they are to follow others \cite{helbing2000simulating}.

While panicked people may choose to follow others, less panicked people may see alternative exits and decide to change direction also.
The probability of such a change in heading per second, of person $i$ towards exit $d$, is given by
\begin{equation*}
P_{i, d}^{d} = \left(1 - \text{e}^{-t_c/t_x}\right) H\left(1 - \delta_{id}\frac{\rho_{d}}{\rho_{m}}\right) \delta_{id}  \left[1 - \exp\left(- \frac{\beta_{id}}{r_{id}^2} \right) \right] (1 - p_i)
\end{equation*}
The first two parts of this equation are the same as the equation for changing direction when in a panic; subsequent changes in a short time and changes in direction towards exits that are visibly congested are considered less likely.  
The term $\delta_{id}$ is included here as we assume people will only choose to switch exits if they can see the exit.
Inside the square parenthesis is a term that limits a person changing direction to doors that are close by. 
The range of this interaction is controlled by $\beta_{id}$ and decays with the distance between the person and the door, $r_{id}$.
The last part of this equation is $(1 - p_i)$ as panicked people are assumed to be less likely to rationally decide to change direction.
The potential exit that a person has the greatest probability of switching to is the only exit they may switch to. This occurs if a uniformly distributed random number is less than the assigned probability.

The magnitude of the desired velocity of a person depends on the level of panic \cite{helbing2005self}, and increases as the person becomes more panicked.
\begin{equation*}
v_i^0 = \left[v_i^{min} + p_i (v_i^{max} - v_i^{min} )\right]
\end{equation*}
The speed linearly increases from walking speed to running speed as panic levels increase from zero to one. In the current model, the maximum speed is randomly assigned according to $v_i^{max} =  v_i^{min} (1 + \text{RND}[0:1] v_f)$, where $\text{RND}[0:1]$ is a random number between 0 and 1, and $v_f$ is the factor by which the speed might increase for panicked people.
The typical occupants of a big-box store can vary significantly with respect to age and physical ability and, therefore, for some the maximum velocity may not be significantly greater than their minimum speed, while for others it could be much greater.
The minimum speed is considered the randomly assigned speed of the person when not in a panicked state.
This desired velocity is used in the linear dynamic equation, along with the forces acting on the person, to allow us to capture the motion of people around each other and obstacles, and their tendency to walk at their desired velocity. Both the linear and angular dynamic equations in the system are iteratively updated using the Velocity Verlet algorithm.
We now apply this model to the dynamics of people evacuating from a big box store.

\section*{Results}

The typical variables used in the simulations presented here, unless otherwise stated, are provided in table 1. 
The mass is assigned as a Gaussian distribution around an average, and the area is assumed to be proportional to the mass for walking people. For people in wheelchairs, the area occupied by the wheelchair is taken to be proportional to the mass of the wheelchair alone, and $m_i$ is the randomly distributed masses of the person and wheelchair combined.  The mass of the wheelchair, $m_w$, in the table is the average mass, and the actual mass is assumed to be proportional to the area calculated from the dimensions $L_i$, $W_i$, $l_i = 0.44 L_i$, and $w_i = 0.71 W_i$ that are randomly assigned according to the data provided by Bharathy and D’Souza \cite{bharathy2018revisiting}.
The moment of inertia of a person standing up is proportional to both mass and area (which is itself proportional to mass) and so this varies as mass squared.
Similarly, for people in wheelchairs the moment of inertia for the sitting person is proportional to the mass of the person squared, while that of the wheelchair is proportional to the area.
People are placed initially at random, such that their location does not overlap with walls or other people. 
The panic level is assigned a random value from a Gaussian distribution with a given average and standard deviation.
The initial panic is assumed to depend on the nature of the emergency and the personal response to the perceived danger. 
All people in a single group have the same heading and desired velocity (either averaged from all group members or taken as that of the slowest group member).
The value of $f_v = 0.75$ is assigned because under normal conditions this does not cause panic to a single individual that might slow down to move around obstacles and around corners in the current model.
$t_s = \qty{1}{\s}$ is taken to be the time over which one might extrapolate future positions to avoid potential collisions, but $t_s = \qty{2}{\s}$ is assumed for wheelchair users as it has been found that wheelchair users reduce their speed ahead of a congestion more readily than people without physical disabilities \cite{geoerg2022people}.

To probe the effects of both rotating the social forces between people and introducing herding in the current model, before turning our attention to the evacuation from a big box store, we consider the interaction of people walking along a simple \qty{2.6}{\m} wide corridor. 
The corridor is \qty{100}{\m} long, but with periodic boundary conditions, and 50 people are randomly placed and assigned desired velocities to either the left or right directions.
The system is allowed to reach a relatively steady state after \qty{100}{\s} and then the position of people across the corridor is averaged over the next \qty{15}{min}, and over 100 independent simulations.
Fig 4. depicts the effects of varying the angle of rotation of the social force interactions.
Fig. 4a. depicts the probability distribution function of the location of people across the corridor.  
When the rotation matrix on the social force term is removed ($\phi = \qty{0}{\degree}$) then the distribution is bimodal. 
In other words, there is equal probability of people passing on the left or right. 
This can be seen in Fig. 4b. which shows a snapshot of a system with $\phi = \qty{0}{\degree}$. 
Even with the angular dependence and elliptical specifications in the social force model, head-on collisions can still occur. 
That said, the model tends to align people in lanes that do not collide, at least until faster people try to overtake slower people traveling in the same direction. 
Increasing the angle $\phi$ not only causes people to predominantly pass on the right, but also reduces the chance of head-on collisions. 
The probability distribution function for both  $\phi = \qty{15}{\degree}$ and $\phi = \qty{30}{\degree}$ are shown, and the inclusion of the rotation matrix in the model is found to cause people to pass on the right. 
Fig. 4c. depicts a snapshot for a system with $\phi = \qty{30}{\degree}$.
The people are confined in their lanes and pass one another on their right-hand side.
It is found that even if two people approach each other directly, on a head-on collision, a value of $\phi = \qty{30}{\degree}$ will cause their paths to deviate enough to not physically interact.

Fig. 5. depicts the effects of panic and herding in people, again confined to a periodic \qty{2.6}{\m} wide corridor.
The fraction of people moving as part of the dominant herd is shown as a function of time, showing the time frame over which dominant herds emerge in the current system. 
There is a transition from lanes of people moving in either direction, at earlier times, to a system where everyone moves in the same direction, at later times.
At early times the people segregate into lanes, as seen in Fig. 5b., with people passing on the right of one another, but with some people having a higher desired velocity than others.
Faster people can find their speeds limited by slower people in front of them, which especially during an emergency evacuation, can cause panic levels to increase. 
Panicked people have a tendency to follow others in the current model of herding.
As people change direction to follow others, they collide with neighboring people and more people are slowed down by these collisions.
This causes more people to become panicked, and this panic can spread like a contagion through the people.
Fig. 5 c. shows a snapshot of the peak level of panic in this system as people change direction, collide with others, and a dominant herd starts to emerge.
A snapshot of the system at later times, when a dominant herd has emerged, is shown in Fig. 5d. 
The herd encompasses everyone in the system (on account of the periodic boundary conditions applied to the corridor) and lanes form as before, but now the faster moving people are to the right and slower moving people are moving to their left. 
Once everyone is part of the same herd, and moving in their desired direction at their desired speed, their panic levels drop steadily with time.

We now consider the evacuation from a big-box store.
While people closer to an exit might make a decision and move towards that exit sooner \cite{haghani2019simulating}, in the current model we simply assign a random start time according to a Gaussian distribution with an average of \qty{10}{\s} and a standard deviation of \qty{5}{\s}.
Fig. 6a. shows the relative egress time as a function of both the effects of the constant $k_p$, which controls the growth in panic levels, and $v_f$ which dictates the random amount that the desired speed increases by in panicked people. 
The relative egress time is the egress time relative to a system with $k_p = 1$ and $v_f = 1$. 
While the relative egress time drops with increasing $k_p$, the effect is only around several percent and plateaus at larger values of $k_p$.
The effects of $v_f$ are more pronounced with a roughly 30\% decrease is egress time with increasing speed for panicked people.
Recall, when $v_f = 0$ there is no increase in speed for a panicked person, and increasing $v_f$ increases the random amount that a panicked persons speed increases by.
It is worth noting that the degree at which people might be able to increase their speed when panicked will depend on a number of factors, but mainly the athleticism of the individuals.
A big-box retail store would not necessarily be associated with athleticism, although this might depend on the time of day (older people shopping on a Tuesday morning versus teenagers out on a Friday evening) and location (a store near a large college or university might be populated with younger people). 
The effects of varying the constants $\beta_v$, $\beta_a$ and $\beta_s$ are depicted in Fig. 6.b. These constants dictate the growth of panic due to visual cues (seeing panicked people nearby), audio cues (hearing panicked people nearby), and the frustration of having ones velocity limited by other people, respectively. 
The relative egress time (relative to a system with $\beta_v = \beta_a = \beta_s = 0.1$) is not sensitive to these parameters for the systems considered here, which suggests that the degree at which panic spreads is not limited to a single mechanism but that all mechanisms that increase panic are playing a part in these systems.
Fig. 6c. shows a snapshot of a system with people evacuating along various paths towards the different exits in the store. 
Here, $k_p = 1$, $v_f = 1$, and $\beta_v = \beta_a = \beta_s = 0.1$. 
In the current model there is an initial level of panic assigned to all people; a Gaussian distribution of mean 0.5 and standard deviation of 0.25 (with $0 \leq p_i \leq 1$ enforced). 
Areas emerge in the simulation where congestion might cause a reduction in speed for some individuals, and the close proximity of people can cause the emergence of elevated panic amongst the evacuees.
These areas are usually around the exits.

The effects of limiting the exits to only the two main entrances at the front of the building is considered in Fig. 7.
Snapshots after \qty{30}{\s} are shown for a system where a) all seven potential exits are open, and b) only the two main entrances are considered as potential exits by the people in the store.
As people funnel towards the two main entrances in the second system we can see more areas of panic emerging due to the increase in congestion.
Fig. 7c depicts the probability of a person evacuating with a given egress time from these systems (statistics accumulated over 100 independent simulations).
Despite the increase in panic (and, hence, velocity) when people are limited to only the two main entrances as potential exits, the egress time increases by on average \qty{7.5}{\s}.
People exiting from the main two exits find it to be less congested when alternative exits are considered by others, while the alternative exits afford people initially located closer to these exits a shorted path to travel. 
The availability of alternative exits, therefore, plays an important role in these evacuation dynamics.

To probe the effects of exit choice further, we plot  the probability of someone initially heading towards a given exit in Fig. 8 (note in the current model people can change their mind, either by following others in a panic or seeing alternative exits that appear more desirable).
Fig. 8a. depicts contour plots of the relative probability of a person choosing an exit in isolation from the other potential exits. 
In particular, the relative probabilities are given by
\begin{equation*}
P_d = P_{d0} \text{e}^{-r_{id}/w_d}
\end{equation*}
where $P_{d0}$ and $w_d$ are constants that control the probability of choosing an exit, and the range over which the exit might be chosen. $r_{id}$ is the initial distance of a person from the door. 
The actual probabilities are normalized to 1 from these relative probabilities. 
In the current model, $P_{d0}$ is equal to 1 for the two main entrances and 0.2 for the remaining 5 exits, and $w_d = \qty{40}{\m}$ for all exits.
Note that this means the main entrances still dominate as the main exits in the model, with roughly 60\% of people choosing the main entrances as their initial exit choice (consistent with experimental observations \cite{shi2009developing}).
However, for people further away from the main entrances, they no longer necessarily have to navigate their way across the store but can exit from much closer locations.
Fig. 8.b. depicts the probability of someone exiting from one of the seven exits with a given egress time. 
Doors 3, 4, 5, and 6 are more isolated from other exits, and especially from the two main exits at the front of the store. 
Therefore, for the (smaller number of) people randomly positioned closer to these exits they might be initially assigned to the closest exit and leave from this exit with a shorter egress time.

The effects of people evacuating in groups is considered in Fig. 9. 
The percentage of people that might belong to a group will depend on the location and time of day (for example, people are more likely to be with friends or family on a weekend \cite{xu2009simulation}). 
The effects of increasing the percentage of people that belong to triads (groups of three people) and dyads (pairs of people) on the relative egress time is depicted in Fig. 9a. 
Here, to isolate the effects of group interactions, the desired speed of all people in a group is taken as the average of all desired speeds of the members of the group.
In other words, while the speed of a group might be limited by the slowest member of the group, here we assign an average velocity to isolate other effects of the group dynamics.
Interestingly, when assigning an average velocity, we find that groups have slightly shorter egress times than individuals.
Figs. 3b thru 3e show snapshots from a simulation where individuals (25\%) are colored black, dyads (50\%) are colored red, and triads (25\%) are colored blue; numbers consistent with what has been found experimentally \cite{challenger2009understanding, xie2020evacuation, corbetta2023physics}.
A particular triad in the center is highlighted as they move through the system.
There are less collisions and interactions with other individuals for each group member, because each group member is surrounded by two other people as part of the same group moving in the same direction and with the same speed.
This is only the case in the center of the store where people in the current model might be initially moving in different directions past one another towards different exits.
Closer to an exit, everyone (regardless of groups) might be expected to be moving in the same direction, albeit with different speeds.
We assume the speed of all group members is the same.
Furthermore, that slower moving individuals are able to increase their speed as much as faster moving individuals are able to reduce their speed. 
Whether or not people in a group match speeds is often based on the social dynamics of the group.
In reality slower moving individuals might be physically incapable to increasing their speeds and groups might be limited by this smaller velocity.
For example, in the simulation with 50\% of the people in dyads and 25\% of the people in triads, limiting the speed of the group to that of the slowest member increases the egress time for people in dyads (on average from \qty{39.14 (1.76)}{\s} to \qty{45.37 (2.41)}{\s}), and people in triads (on average from \qty{38.69 (2.97)}{\s} to \qty{46.52 (4.04)}{\s}). 
Interestingly, individuals surrounded by slower groups can see their egress times increase from, on average, \qty{40.83 (2.18)}{\s} to \qty{44.96 (2.69)}{\s}.
In other words, people in groups may take longer to evacuate depending on how the desired velocity of the group is determined, and this can influence egress times for individuals that are not part of a group.
In the current simulation we also assume the time to start moving for groups is no longer than for individuals, which might not be the case when groups have to confer before reaching a decision \cite{bode2015disentangling}.
This might further increase the egress time for groups, depending on the specifics of their social dynamics.

The effects of increasing the number of people on crutches and people using wheelchairs is depicted in Fig. 10.
The percentage of people leaving as a function of egress time is plotted in Fig. 10a.
People without disabilities, people who use crutches and people who use wheelchairs, in a system with both 2\% of crutch users and 2\% wheelchair users, are contrasted with people without disabilities in a system where no one is disabled. 
The speed of people with crutches (\qty{0.94 (0.30)}{\m\per\s}) and people using wheelchairs (\qty{0.79(0.35)}{\m\per\s}) is on average lower than the speed of people without a disability (\qty{1.36(0.26)}{\m\per\s}) \cite{hashemi2018emergency}.
This is reflected in the egress times.
People without a disability have a shorter egress time of \qty{39.82 (1.51)}{\s} on average. 
In contrast, people that use crutches have an egress time of \qty{45.39 (7.89)}{\s} on average, while people that use wheelchairs have an egress time of \qty{65.84 (32.97)}{\s} on average.
The effects of people using either crutches or wheelchairs on the egress time of people without a disability is negligible in the current system; 
people without a disability in systems without people using crutches or wheelchairs have an egress time of \qty{39.36 (1.62)}{\s} on average. 
Even in systems where the evacuation is limited to the main two exits, and congestion is expected to play a larger role, the average egress time only increases for people without disabilities from \qty{47.42 (0.93)}{\s} to \qty{47.86 (1.00)}{\s}, in systems with 2\% of people using crutches and 2\% of people using wheelchairs.
Fig. 10b depicts the relative egress time as a function of the percentage of people using wheelchairs, and as a function of the percentage of people using crutches. 
Again, the changes in egress time for the systems considered here do not significantly depend on the number of people using either crutches or wheelchairs.
The number of people using crutches or wheelchairs are expected to be small.
The main exits in the large box store are also large and the occupancy is limited; in other scenarios, where congestion and trampling might occur, the physical dimensions of the wheelchairs may be more detrimental to egress times, especially if people are less able to move around slower moving wheelchairs. 

\section*{Conclusions}

A social force model of evacuation dynamics has been developed that captures the tendency of people to pass one another on a given side, accounts for the non-circular cross-sectional area of people, and incorporates irregular shapes that capture the physical dimensions of people using wheelchairs.
The model further captured the spread of panic within a crowd and associated herding behavior.
It was found that, when applied to a typical big box store, that limiting the perceived availability of exits can significantly increase egress times, especially for people that might be further from the main entrances when the evacuation occurs.
Future work will apply the model to more aggressive scenarios. Evacuations involving elevated congestion and crushing, that will result in injury, people falling to the ground or wheelchairs being pushed over. This will involve the inclusion of immobile people, along with mutual aid behavior as evacuees see others in trouble and move to assist them.

\bibliographystyle{unsrt} 
\bibliography{paper.bib}

\clearpage

\begin{center}
\includegraphics[width=0.8\linewidth]{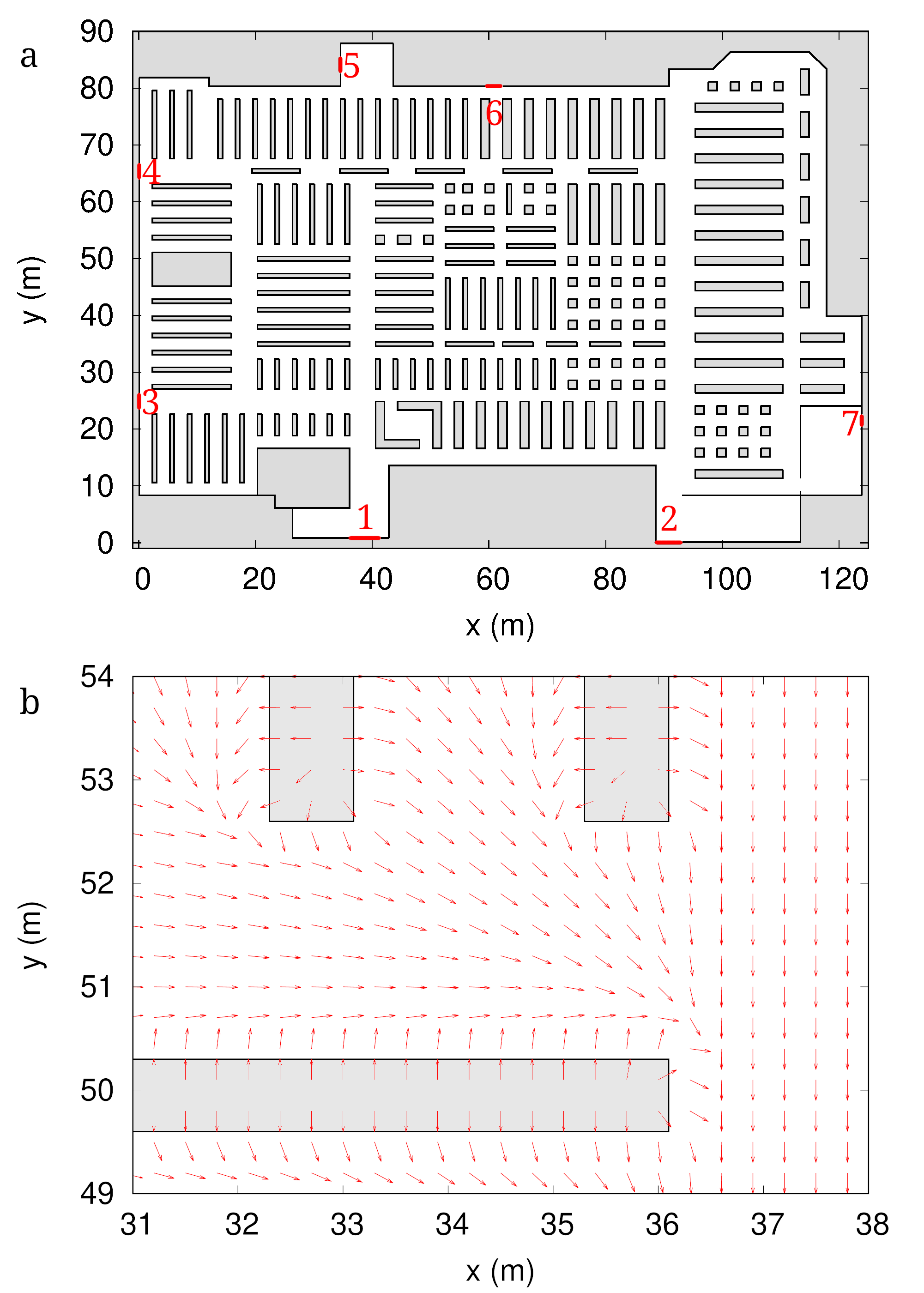}
\end{center}

Figure 1: Idealized layout of a typical large big-box store in the United States. a) The layout with shelving and other physical structures shaded gray, and the aisles and other free space depicted as white space. The seven potential exits are numbered. b) A close up of the system depicting the direction, $\mathbf{\hat{e}_i}$, that optimizes the distance to a given exit. This unit vector field is calculated with a spatial resolution of \qty{10}{\cm}.

\clearpage

\begin{center}
\includegraphics[width=\linewidth]{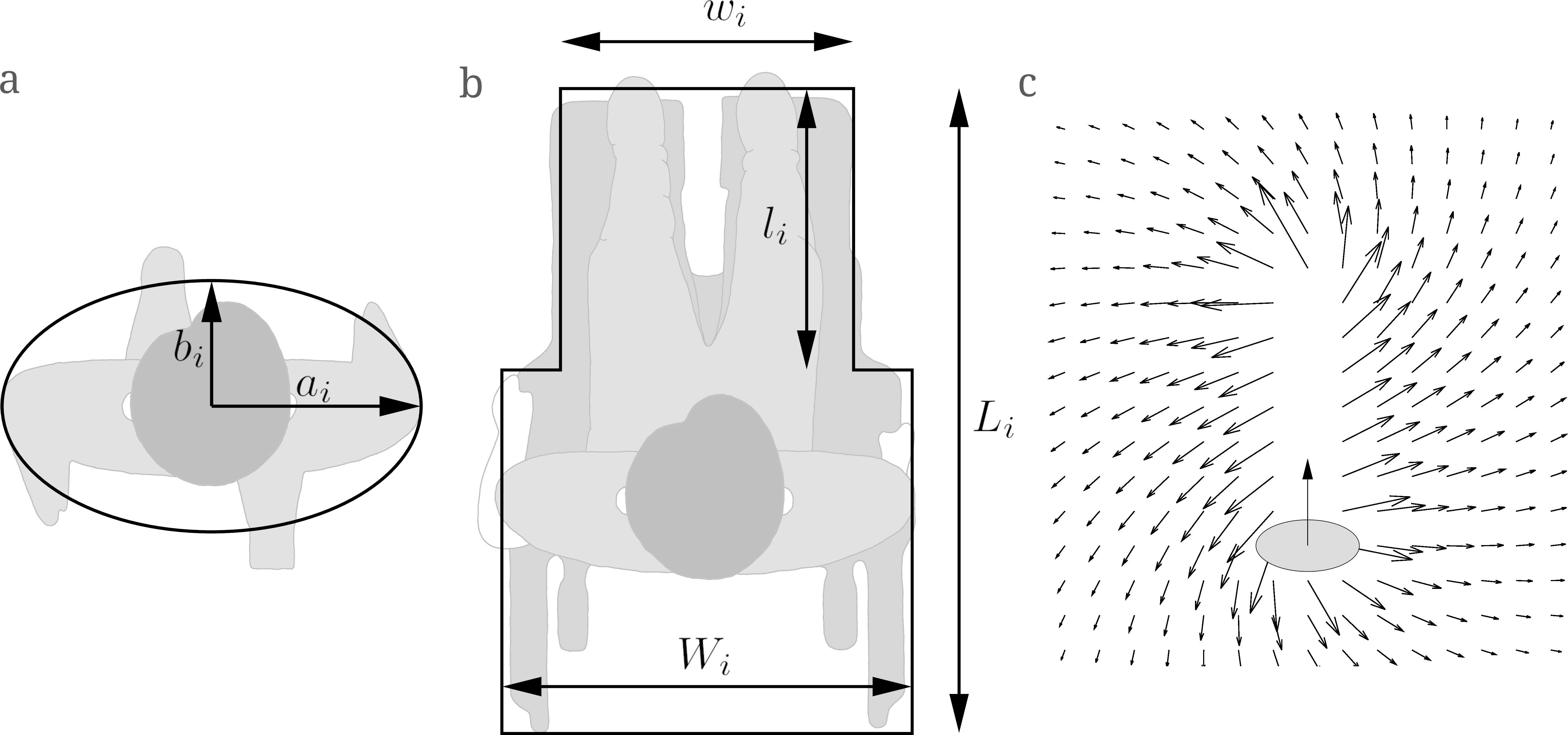}
\end{center}

Figure 2: The simulation of people in the current social force model. a) Able-bodied people and people with crutches are physically approximated by ellipses. b) Wheelchair-users are physically approximated by an irregular polygon. c) Socially, the repulsive force acting on a neighbor extends along the direction of motion of a person. This force can be rotated to mimic the tendency of people to pass on a given side when walking past each other in opposite directions.

\clearpage

\begin{center}
\includegraphics[width=\linewidth]{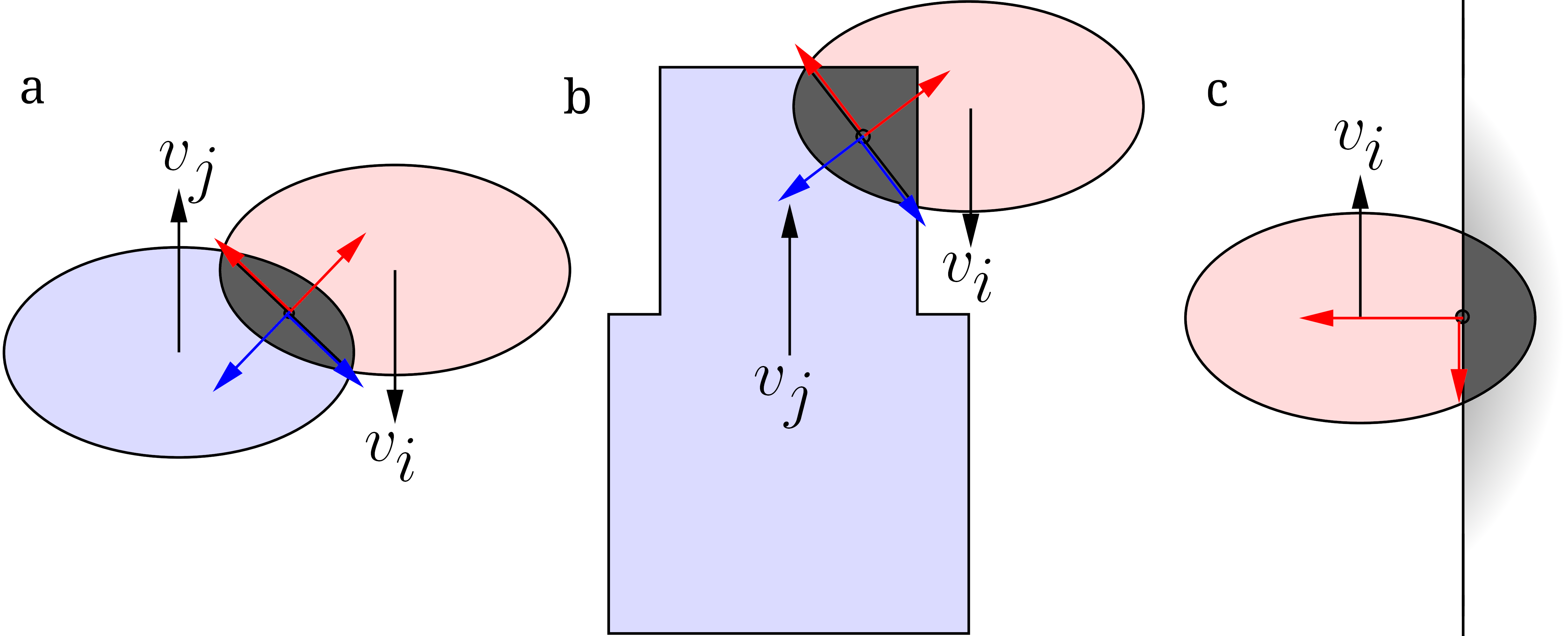}
\end{center}

Figure 3: The physical interaction between a) two people without disabilities (or people with crutches), b) a person not using a wheelchair and a person using a wheelchair, and c) a person without a disability (or person with crutches) and a wall. The magnitude of the interaction is proportional to the area of overlap. A normal force acts perpendicular to the line passing through the intersection points of the overlapping shapes and a tangential forces acts parallel to the line passing through the intersection points and is proportional to their relative velocity in the tangential direction.

\clearpage

\begin{center}
\includegraphics[width=0.8\linewidth]{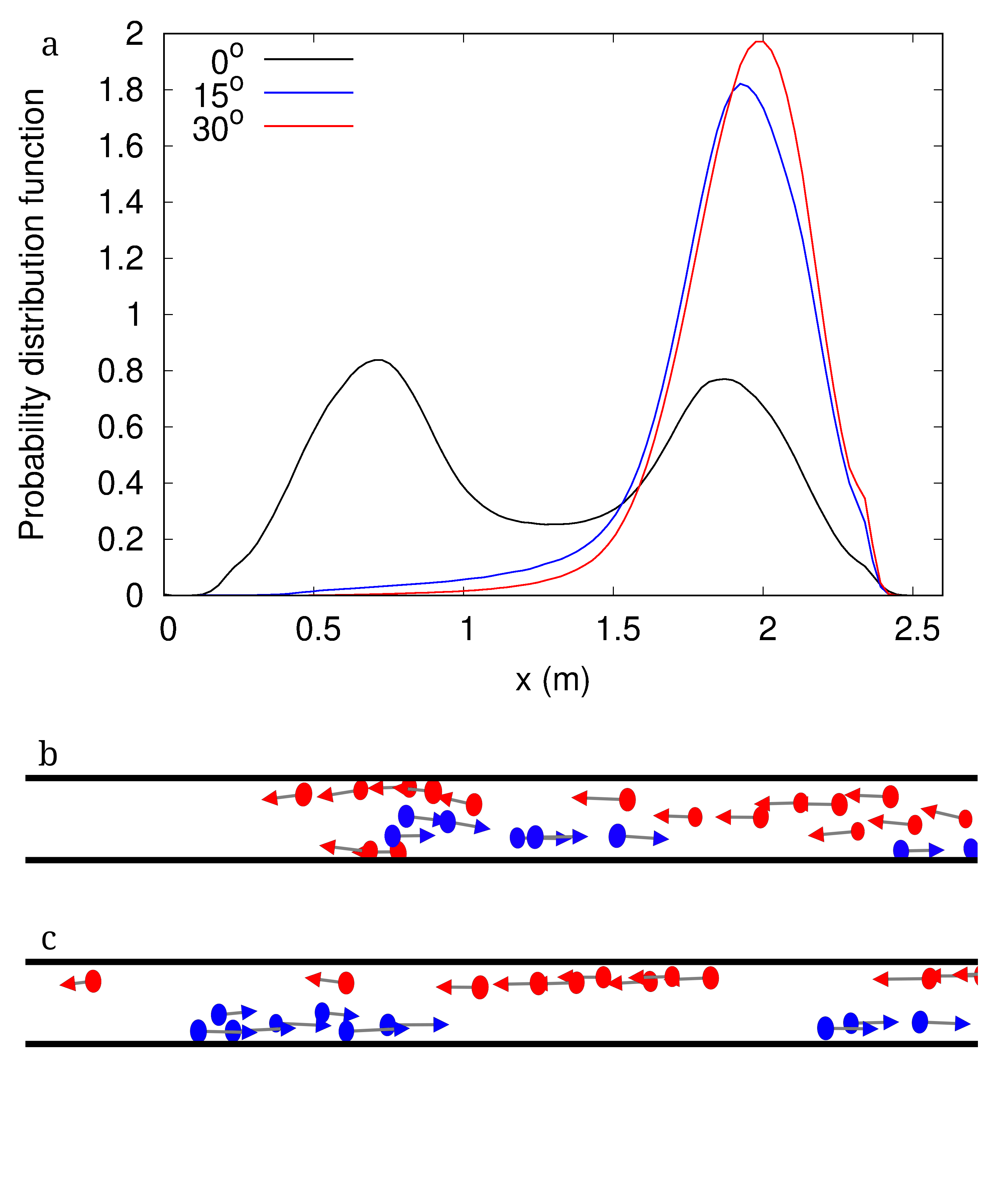}
\end{center}

Figure 4: A simple model of a corridor (width of \qty{2.6}{\m}) with a variety of people moving in either direction. a) The probability of finding a person at a location across the width of the corridor as a function of the angle, $\phi$, used in the rotation matrix, $\mathbf{R}_{\phi}()$, acting on the social force. Snapshot of the system with b) $\phi = \qty{0}{\degree}$ and c) $\phi = \qty{30}{\degree}$ after several seconds of movement. People moving left are colored red and people moving right are colored blue. While lanes are created in either scenario, people pass on the right when $\phi = \qty{30}{\degree}$.

\clearpage

\begin{center}
\includegraphics[width=0.75\linewidth]{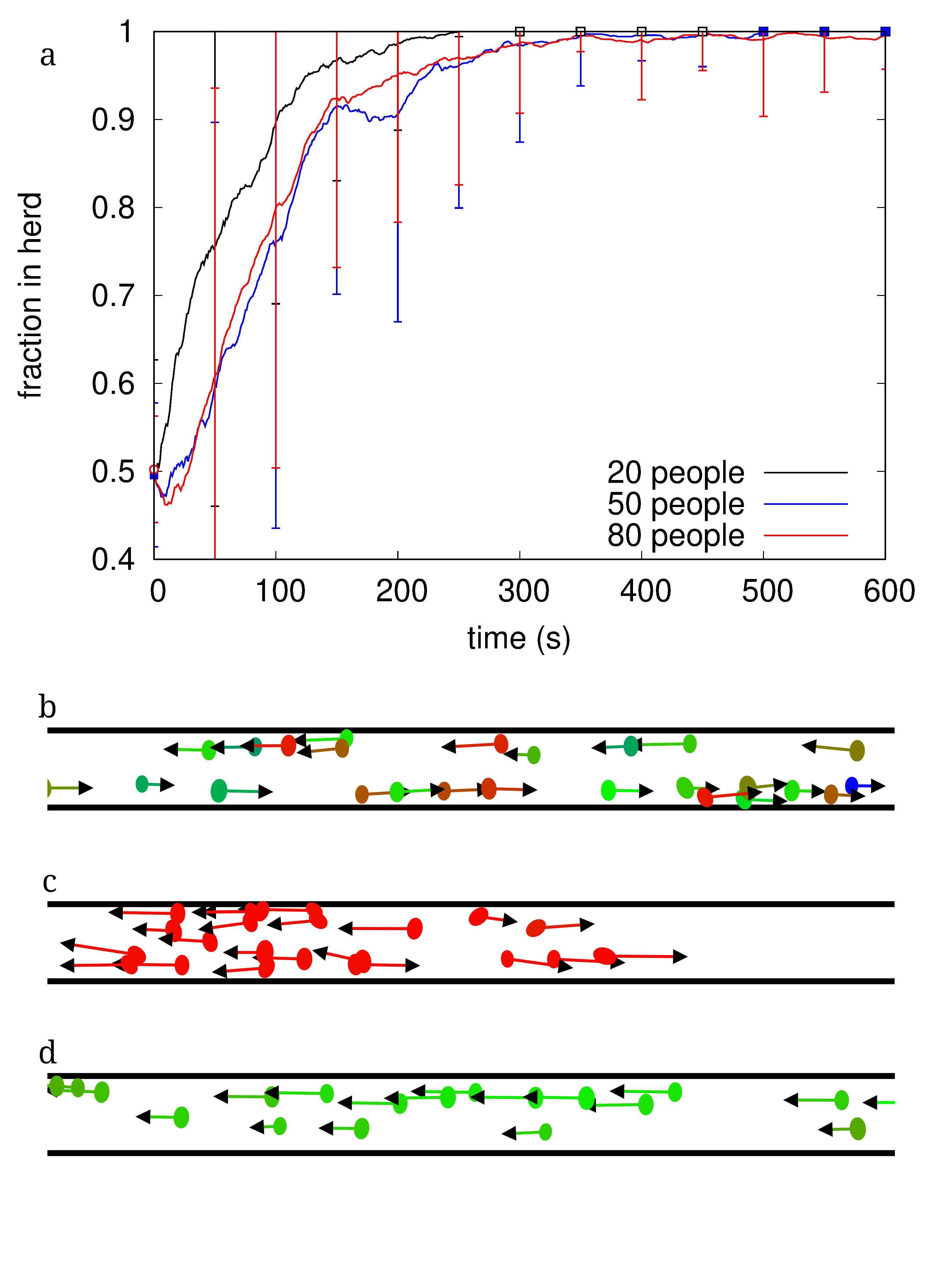}
\end{center}

Figure 5: Herding of people in a simple corridor (width of \qty{2.6}{\m}) that consists of people either heading left or right. a) The fraction of people that are in the dominate herd versus time, as a function of the density (number of people in a \qty{100}{\m} long corridor with periodic boundary conditions). Snapshot of a portion of a system with 80 people at b) \qty{100}{\s}, c) \qty{220}{\s}, and d) \qty{640}{\s}. The color of the people varies with panic from blue ($p_i = 0$), through green  ($p_i = 0.5$), to red ($p_i = 1$); the initial panic is $p_i = 0.5 \pm 0.25$. Initially, people become panicked as they can't pass slower neighbors. During intermediate times, as people collide, they become increasingly panicked. At late times, once everyone is in the same herd and heading in the same direction the panic levels steadily drop.

\clearpage

\begin{center}
\includegraphics[width=0.5\linewidth]{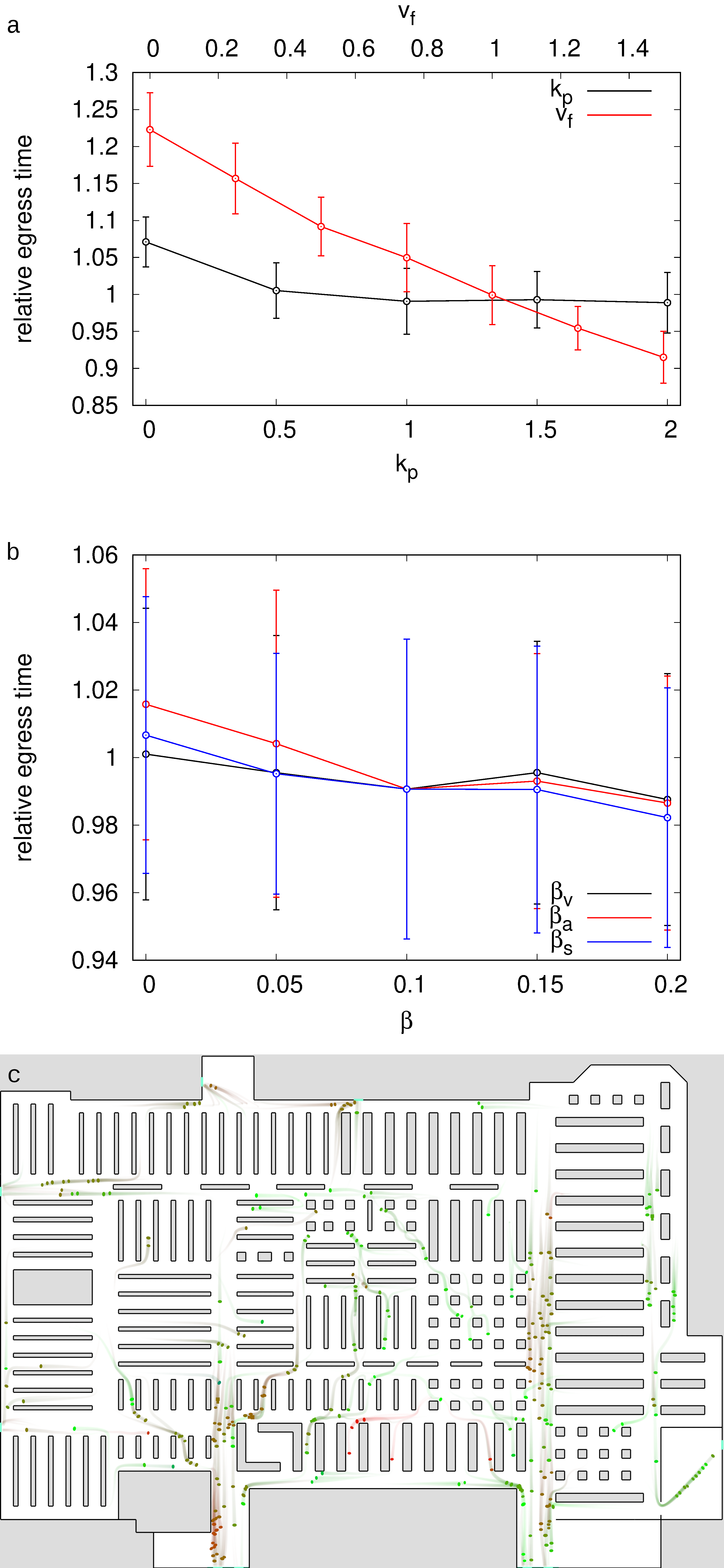}
\end{center}

Figure 6: The effects of panic in the evacuation from the big-box store. a) Increasing $k_p$, the constant in the panic evolution equation, and $v_f$, the factor that the velocity might increase for panicked people, decreases the egress time (presented relative to egress time when $k_p = 1$ and $v_f = 1$). b) The relative egress time as a function of the $\beta$ constants. c) A snapshot of a simulation with $k_p = 1$ and $v_f = 1$ where the color represents the degree of panic from blue ($p_i = 0$), through green  ($p_i = 0.5$), to red ($p_i = 1$); the initial panic is $p_i = 0.5 \pm 0.25$. Trails are shown to depict the path of the people, and areas where people interact more are shown to exhibit more panicked behavior.

\clearpage

\begin{center}
\includegraphics[width=0.5\linewidth]{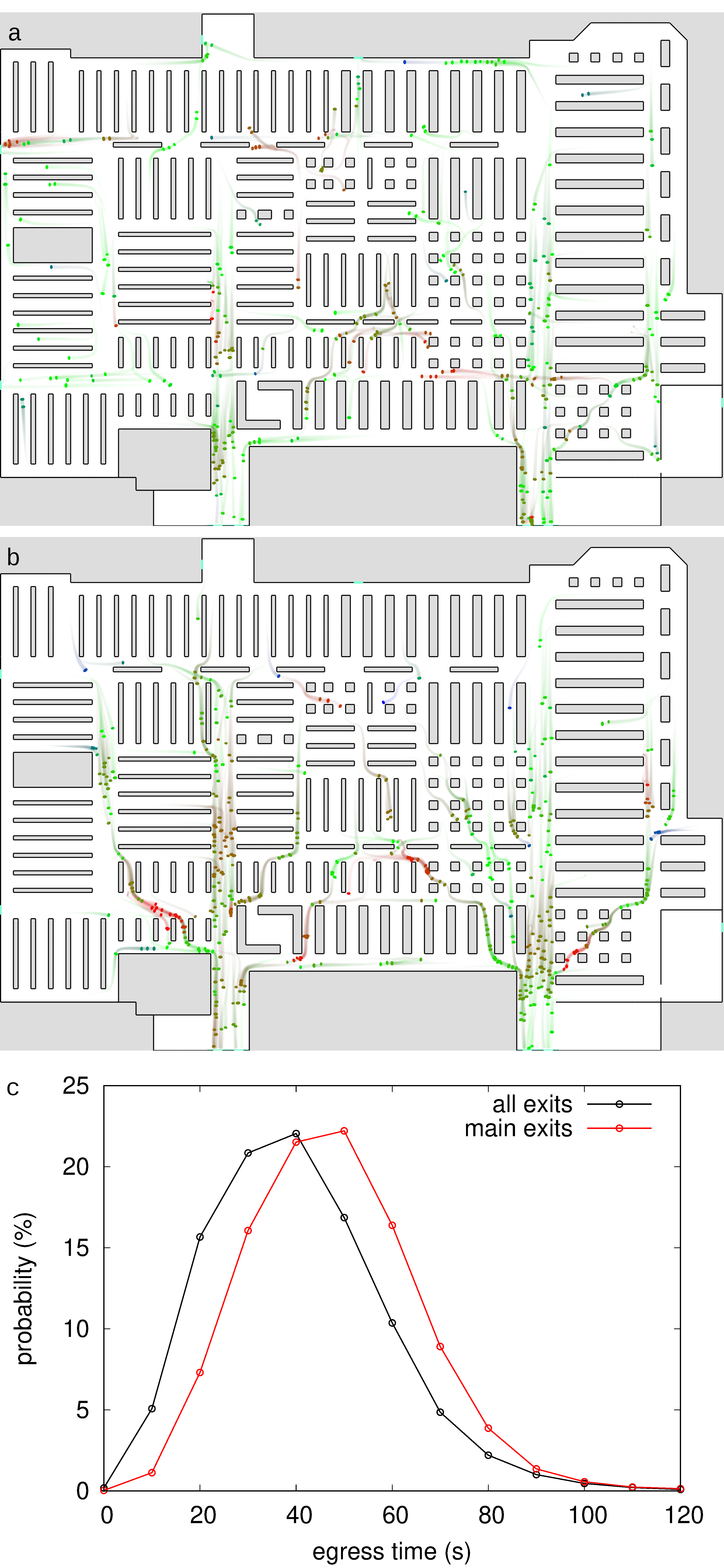}
\end{center}

Figure 7: The effect of exit choice. a) A snapshot of a simulation (after \qty{30}{\s}) when all exits are considered and b) a snapshot of a simulation (after \qty{30}{\s}) when only the main exits (at the front of the store) are considered. c) The probability of a person having a given egress time is plotted for when all exits are considered and when only the main exits are considered. The average time can be reduced by on average \qty{7.5}{\s} when considering all exits.

\clearpage

\begin{center}
\includegraphics[width=0.8\linewidth]{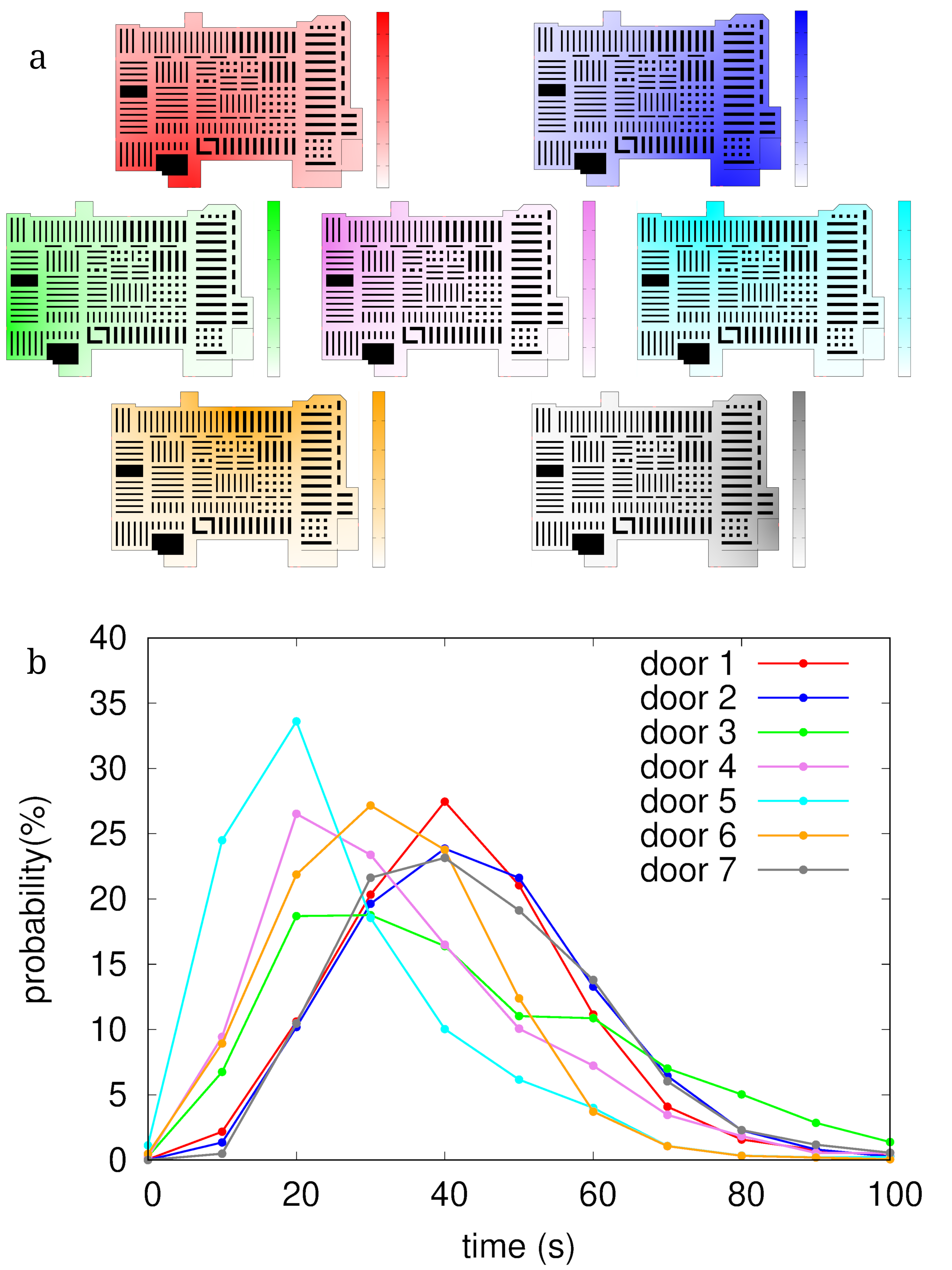}
\end{center}

Figure 8: The effects of exit choice. a) The probability distributions for each exit are shown; exits are colored in sequence corresponding to the labeling in b. b) the probability of a person exiting an exit at a given time. Note the egress times from doors 3, 4, 5 and 6 are lower as these exits are further from the main exits and only a small number of people close to these exits are likely to choose them.

\clearpage

\begin{center}
\includegraphics[width=0.6\linewidth]{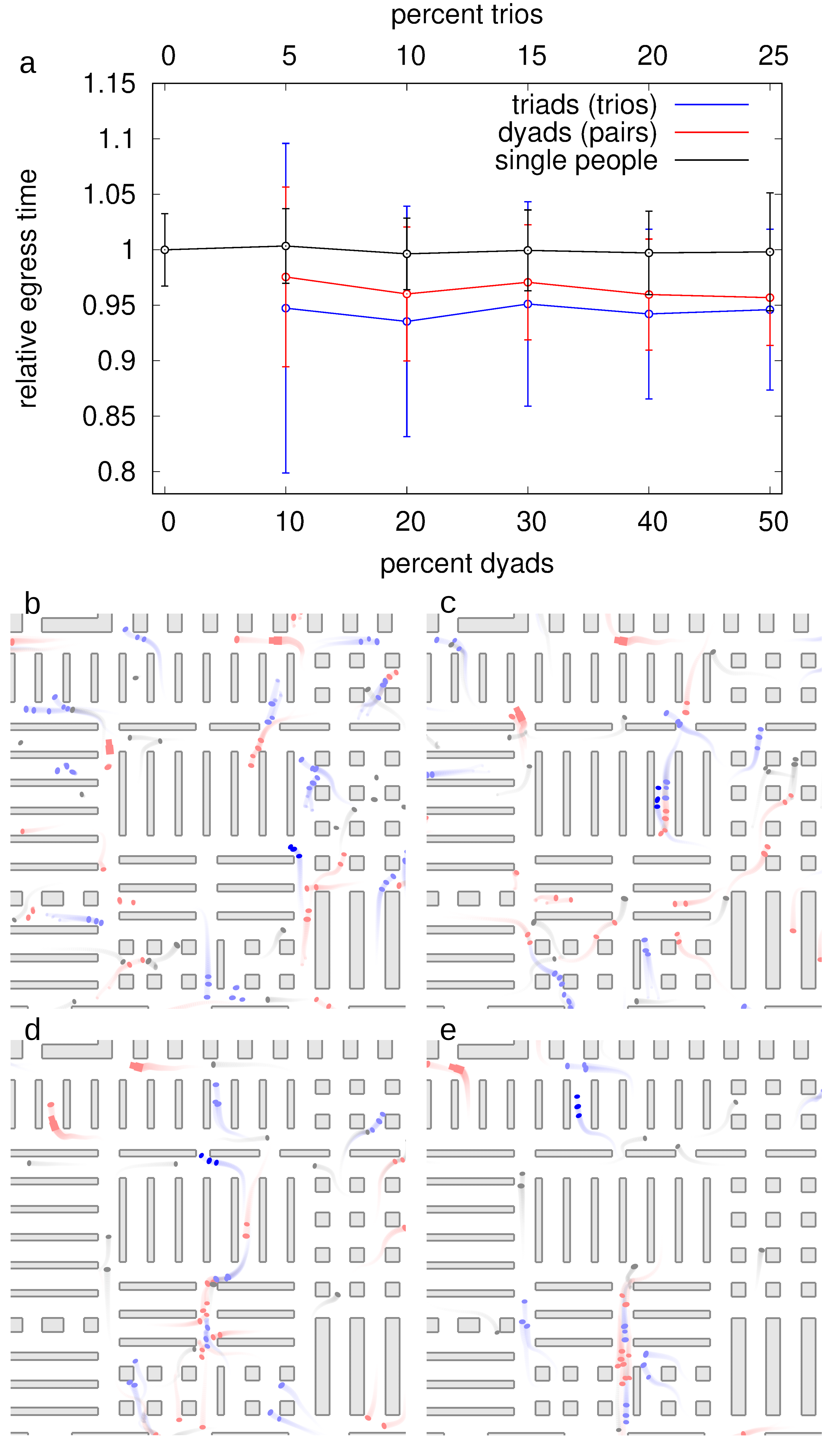}
\end{center}

Figure 9: The effects of dyads and triads on the relative egress time. a) The relative egress times for single people, dyads and triads as the number of triads increases to 25\% and the number of dyads increases to 50\% of the population. Snapshots of a system with single people (black), dyads (red) and triads (blue) at b) $t = \qty{15}{\s}$, c) $t = \qty{20}{\s}$, d) $t = \qty{25}{\s}$, and e) $t = \qty{30}{\s}$. A triad group moving up in the middle of the system is highlighted.

\clearpage

\begin{center}
\includegraphics[width=0.8\linewidth]{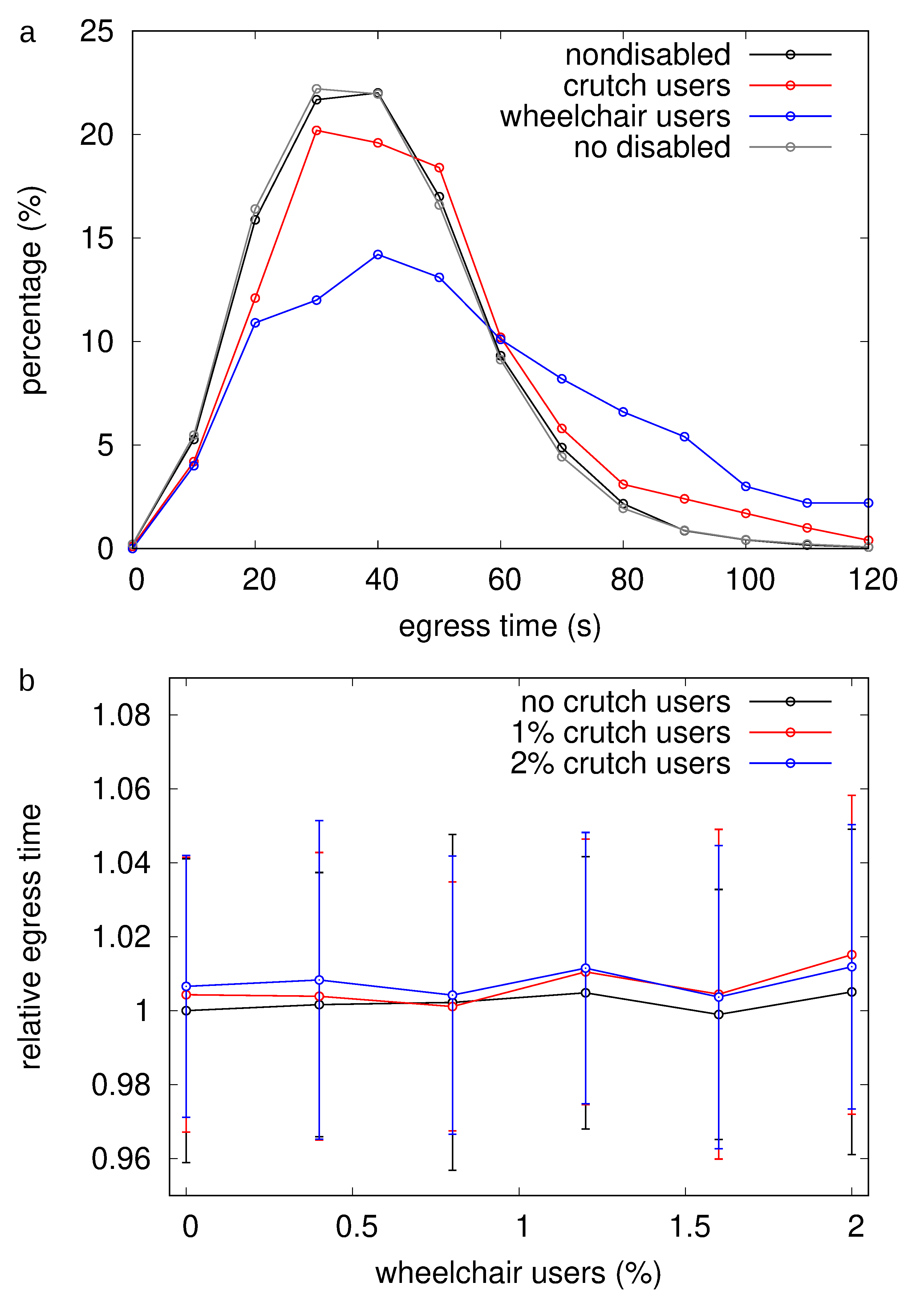}
\end{center}

Figure 10: The effects of people with crutches and people who use wheelchairs on egress times. a) The percentage of people without a disability, people with crutches and people using wheelchairs leaving as a function of egress time. A system with both 2\% of people using crutches and 2\% of people using wheelchairs is contrasted with a system with only people without disabilities. b) The relative egress time as a function of the percentage of people using wheelchairs is depicted in systems with different percentages of people using crutches.

\clearpage

\begin{table}[H]
\begin{center}
\begin{tabular}{|c|c|l|}
\hline
variable & value &  \\ \hline
$m_i$ & \qty{84.05(20.35)}{\kg}  &  \cite{fryar2021anthropometric}\\
$m_w$ & \qty{12}{\kg}  &  \cite{fu2024application}\\
$v_i^0$ & \qty{1.36(0.26)}{\m\per\s} & \cite{helbing1995social} \\
$v_i^0$ & \qty{0.94 (0.30)}{\m\per\s} (crutches) & \cite{hashemi2018emergency} \\
$v_i^0$ & \qty{0.79(0.35)}{\m\per\s} (wheelchair) & \cite{hashemi2018emergency} \\
$t_v$ & 0.5 &  \\
$\lambda$ & 0.1 & \cite{helbing2010pedestrian} \\
$A_i$ & \qty{20}{\N} &  \\
$A_i$ & \qty{80}{\N} (wheelchair) &  \\
$A_i$ & \qty{10}{\N} (group) &  \\
$B_i$ & \qty{1}{\m} &  \\
$t_s$ & \qty{1}{\s} &  \\
$t_s$ & \qty{2}{\s} (wheelchair) &  \\
$\phi$ & \qty{30}{\degree} &  \\
$k_n$ & \qty{1.2e5}{\N\per\m}  & \cite{helbing2000simulating} \\
$k_n$ & \qty{1.2e7}{\N\per\m} (wheelchair)  &  \\
$k_t$ & \qty{2.4e5}{\N\per\m}  & \cite{helbing2000simulating} \\
$k_g$ & \qty{20}{\N\per\m} &  \\
$d_0$ & \qty{0.25}{\m} (groupof 2) &  \\
$d_0$ & \qty{0.5}{\m} (groupof 3) &  \\
$d_0$ & \qty{2}{\m} (wheelchair) &  \\
$A_w$ & \qty{200}{\N}  &  \\
$A_w$ & \qty{400}{\N}  (wheelchair)  &  \\
$B_w$ & \qty{0.25}{\m} &  \\ \hline
  \end{tabular}
  \hspace{1cm} 
\begin{tabular}{|c|c|l|}
 \hline
variable & value &  \\ \hline
$\mu_s$ & 0.9 &  \\
$\mu_k$ & 0.7 &  \\
$I_i$ & \qty{1.28}{\kg\square\m} & \cite{santschi1963moments} \\
$I_i$ & \qty{3.78}{\kg\square\m} (sitting) &  \cite{santschi1963moments} \\
$I_i$ & \qty{5.23}{\kg\square\m} (wheelchair) &  \cite{wang2007experimental}\\
$k_{\theta}$ & \qty{4}{\N\m}  &  \\
$k_{\theta}$ & \qty{200}{\N\m} (wheelchair) &  \\
$a_i$ & \qty{0.25}{\m} & \cite{fu2024application} \\
$a_i$ & \qty{0.3}{\m} (crutches) & \cite{fu2024application} \\
$b_i$ & \qty{0.15}{\m}  & \cite{fu2024application} \\
$L_i$ & \qty{1.169(0.131)}{\m}  & \cite{bharathy2018revisiting} \\
$W_i$ & \qty{0.694(0.066)}{\m}  & \cite{bharathy2018revisiting} \\
$l_i$ & $0.44 L_i$  &  \\
$w_i$ & $0.71 W_i$  &  \\
$\beta_v$ & 0.1 &  \\
$\beta_a$ & 0.1 &  \\
$\delta_i$ & 0.01 &  \\
$\beta_s$ & 1. &  \\
$f_v$ & 0.75 &  \\
$t_x$ & \qty{20}{\s} &  \\
$\rho_m$ & \qty{4}{people \per\square\m} & \cite{imo2002interim}\\
$\beta_{ijd}$ & \qty{3}{\square\m} &  \\
$\beta_{id}$ & \qty{20}{\square\m} &  \\ \hline
\end{tabular}
\end{center}
\end{table}

\normalsize
Table 1. Parameters assumed in the current model and sources when applicable.

\end{document}